# Quantitative Measurements of Electromechanical Response with a Metrological Atomic Force Microscope

Aleksander Labuda and Roger Proksch
*Asylum Research an Oxford Instruments Company, Santa Barbara, CA, USA, 93117.*

An ongoing challenge in atomic force microscope (AFM) experiments is the quantitative measurement of cantilever motion. The vast majority of AFMs use the optical beam deflection (OBD) method to infer the deflection of the cantilever. The OBD method is easy to implement, has impressive noise performance and tends to be mechanically robust. However, it represents an indirect measurement of the cantilever displacement, since it is fundamentally an angular rather than a displacement measurement. Here, we demonstrate a metrological AFM that combines an OBD sensor with a laser Doppler vibrometer (LDV) to enable accurate measurements of the cantilever velocity and displacement. The OBD/LDV AFM allows a host of quantitative measurements to be performed, including *in-situ* measurements of cantilever oscillation modes in piezoresponse force microscopy (PFM). As an example application, we demonstrate how this instrument can be used for accurate quantification of piezoelectric sensitivity – a longstanding goal in the electromechanical community.

Since its invention, much of the effort invested in research with the atomic force microscope[1] (AFM) has implicitly or explicitly involved interpreting the measured cantilever motion in terms of interactions between its tip and the sample. Although some AFMs have employed interferometric detection schemes[2] the most common method for measuring this motion is the optical beam deflection (OBD) method[3,4], also known as the "beam bounce" method. The OBD method uses the reflected angle of a laser focused on the back of the cantilever to determine the cantilever tip position. This method measures the *angular* changes of the cantilever, rather than the *displacement* of the tip. Measuring the angular deflection of the cantilever requires additional interpretation to relate the measurement to tip-sample interactions. In particular, assumptions about the cantilever mode shape are required to relate the measured angle to the displacement of the tip. These assumptions often fail, especially when the cantilever tip is in contact with the surface.

To date, OBD and interferometry have typically been discussed as interchangeable methods of detection and differentiated mostly on a technical level in terms of instrumental implementation. Here, we demonstrate that the differences between these methods are fundamental, with each method providing complementary information about tip-sample interactions.

This letter describes a so-called metrological AFM that combines a conventional OBD sensor with a laser Doppler vibrometer[5] (LDV), which interferometrically measures the velocity of an object from the Doppler shift of a reflected laser beam. A key advantage of interferometric methods is that the sensitivity is intrinsically and accurately calibrated, since the calibration is based on the well-defined wavelength of light. Furthermore, interferometry measures the tip velocity (or displacement) directly and therefore requires no assumptions about the cantilever mode shape, as long as the laser spot is directly above the cantilever tip. This combined OBD/LDV AFM allows both detection methods to be used simultaneously, enabling LDV mapping of cantilever (or sample) motion at any user-selected position in the optical view during regular AFM experiments performed with OBD.

Here, the capabilities of the combined OBD/LDV AFM are demonstrated in the context of piezoresponse force microscopy[6] (PFM). PFM is based on the converse piezoelectric effect. After putting the cantilever tip in contact with a piezoelectric sample, the tip-sample bias voltage is modulated periodically. This generates an oscillating electric field below the tip and leads to localized deformations in the sample surface. The resulting sample vibrations act as a mechanical drive for the cantilever tip. The magnitude of effective piezoelectric response of the surface $d_{eff}$, in pm/V, is measured as the amplitude of the tip displacement divided by the amplitude of the tip-sample voltage. In addition, the phase of the response provides information about the polarization direction.

Typically, higher frequency PFM measurements allow faster scanning which effectively reduces $1/f$ noise and drift, and are essential for rapid domain mapping. However, it is well known that the drive frequency of the electrical excitation can have a profound effect on the measured signal.[7,8] Since the frequency response of most ferroelectric samples should be flat into the GHz range,[9] this suggests that some features in the frequency response into the MHz range may originate from cantilever dynamics instead of ferroelectric effects.[10,11] In order to minimize the effects of cantilever resonances on the ferroelectric signal, single-frequency PFM has mostly been limited to operation at a few hundred kHz or lower,[12] with some exceptions.[13] Two- or multiple-frequency techniques such as dual AC resonance tracking[14] (DART) and band excitation[15] (BE) have reduced the severity of the problem by tracking the resonance frequency, but to a limited degree.

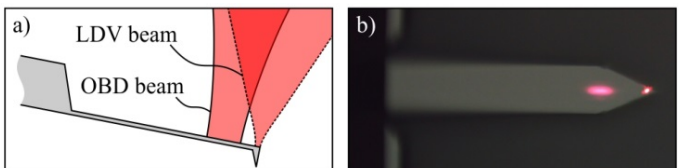

Figure 1: (a) Diagram showing a side view of the optical paths for the LDV and OBD beams focused onto the cantilever. (b) Corresponding photograph of a cantilever from above showing the LDV and OBD spots.

In addition, there are other forces present that respond to tip-sample bias modulation at any drive frequency, such as delocalized electrostatic forces between the body of the cantilever and the sample surface charge.[16,17] In many cases, the undesirable response of the cantilever to these electrostatic forces overwhelms the PFM signal of interest. Over the years, a number of approaches for maximizing the PFM response and minimizing or eliminating the electrostatic components have been developed; however, this issue remains a significant challenge.[18,19] Misinterpreting the electrostatic signal as a tip displacement can lead to incorrect estimation of the piezoelectric sensitivity and relative phase response.

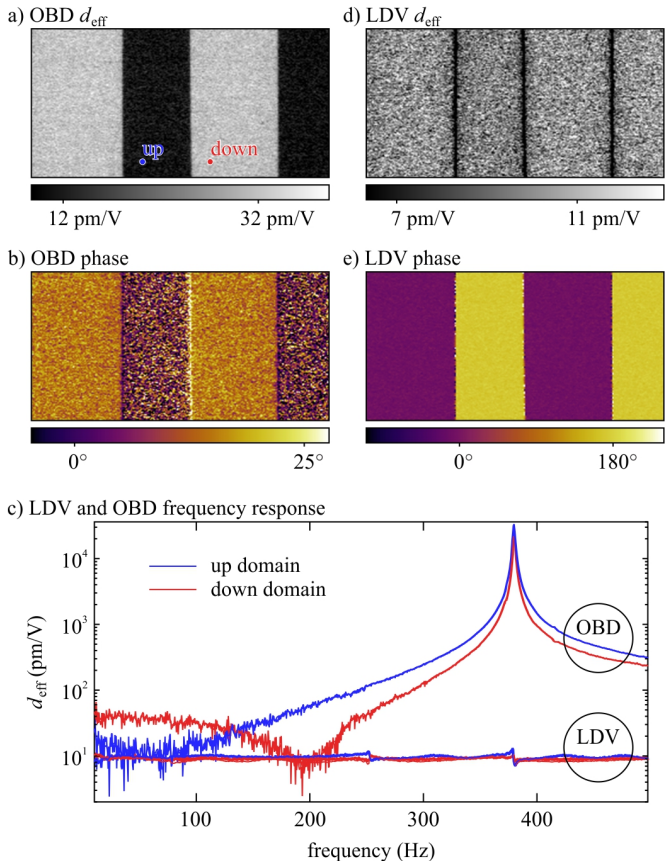

Figure 2: (a,d) OBD and LDV measurements of the effective piezoelectric sensitivity $d_{eff}$ and (b,e) phase over domains in a periodically poled lithium niobate reference sample. Drive frequency: 25 kHz. Scan size 5 μm × 10 μm. (c) For the OBD sensor, the frequency response is dominated by cantilever dynamics, both appearing quite different over oppositely poled domains and varying in magnitude over a factor of 1000×. The measured LDV response is nearly frequency independent. The frequency range for both measurements spans nearly 500 kHz.

The metrological AFM used in this study combines a commercial Cypher AFM (Asylum Research, Santa Barbara, CA) with an integrated quantitative LDV system (Polytec GmbH, Waldbronn, Germany) to achieve highly sensitive electromechanical imaging and spectroscopy. Figure 1 illustrates that the LDV and OBD spots are both focused onto the cantilever. The spots can be separately positioned and focused. By virtue of its large numerical aperture, the LDV spot is focused down to ~2 μm. This allows high-resolution mapping of the cantilever dynamics by local measurements of its displacement. Unlike OBD, LDV sensitivity is not affected by a reduction in spot size. More importantly, because the LDV measurement is encoded as a frequency (Doppler) shift of the helium-neon laser, the sensitivity is highly accurate and does not change with the optical reflectivity of the cantilever nor with laser power.

Periodically poled lithium niobate (PPLN) was chosen as a reference sample for this study due to its availability, independently characterized properties and large domains.[20] Of relevance to this study is that uniaxial PPLN should exhibit the following characteristics in an ideal PFM measurement: (i) frequency-independent response,[21] (ii) amplitude independence of the ferroelectric polarization direction[22] and (iii) 180° phase shift across oppositely polarized domains.

OBD and LVD measurements of the PPLN sample are compared in Fig. 2. In Fig. 2(a), the OBD $d_{eff}$ measurements show significant variations across both poled domains despite the expected amplitude independence. Furthermore, Fig. 2(b) reveals the very low phase contrast between domains of ~20° measured by OBD, well below the expected 180° phase shift. Figure 2(c) shows the frequency-dependent amplitude response over the two domains, similar to that described elsewhere.[23] These results demonstrate the frequency response that is symptomatic of the problems observed with OBD in PFM measurements. The OBD measured response is dominated by information about the cantilever bending, which cannot be easily related to tip motion.

In stark contrast, the LDV measurements show equal amplitudes over both poled domains in Fig. 2(d), with variations approximately an order of magnitude smaller than those of the ODB measurements in Fig. 2(a). In addition, Fig. 2(e) contains the expected 180° phase shift from oppositely poled areas. Figure 2(c) shows that the OBD response varies more than 1000× over a 500 kHz frequency range, consistent with the measurement being dominated by the cantilever dynamics. Note that with respect to the OBD measurement, the LDV response shows very little variation over the entire measured frequency range. There is a small remnant kink in the response at the contact resonance frequency, which is discussed below.

Positioning the LDV spot in different locations on the cantilever relative to the tip location allows direct investigation of the cantilever dynamics that occur in PFM experiments. Fig. 3(a) illustrates three distinct scenarios: the laser spot is located on either side of the tip, or directly above the tip. Fig. 3(b) shows the evolution of the system transfer function as the LDV spot is moved along the length of cantilever. As the laser spot is moved towards the end of the cantilever, an anti-resonance sweeps upward in frequency around the contact resonance peak. When the LDV spot is located immediately above the tip (black curve), the resonance and anti-resonance pair cancels out and leads to a nearly flat response. In this specific location, the LDV signal is blind to the dynamics of the cantilever and reports only the displacement of the tip, as can be understood by inspection of Figure 3(a). This situation is ideal for quantifying surface strain.

Fig. 3(c) demonstrates how the LDV spot location affects the measured response. Although the images were acquired at a drive frequency of 25 kHz, well below the contact resonance frequency of 380 kHz, the cantilever dynamics still have significant impact on the measured values of $d_{eff}$ between different domains. In this scenario, the LDV measurement couples both the tip displacement and the cantilever dynamics. As explained in the previous paragraph, it is only when the laser spot is directly above the tip that the measurement is decoupled from the cantilever dynamics.

In Fig. 3(b), it is important to note that the frequency location of the anti-resonance for a given laser spot location also depends on the polarity of the surface – not only the laser spot position. This leads to the undesirable contrast in the amplitude response on both surfaces seen in Fig. 3(c). Note that similar coupling of cantilever dynamics is the source of spurious contrast in OBD measurements. However, there is no location for the OBD laser spot that eliminates the cantilever dynamics from the measurement of tip displacement.

These results suggest a methodology for accurately quantifying the electromechanical response of a sample. Once tip-sample contact is established with a chosen OBD deflection setpoint, the contact resonance frequency is identified by electrically driving the cantilever. Then, the LDV spot position is optimized by iterative minimization of the measured frequency response around the resonance frequency. Finally, after achieving a flat frequency response around the contact resonance, conventional sub-resonant electromechanical imaging can be performed with much higher accuracy. This protocol greatly extends the available frequency range for accurate PFM measurements, which is now limited only by the precision in positioning the LDV laser spot directly above the cantilever tip.

To demonstrate these ideas, the measurements in Fig. 2 were repeated with five different cantilevers with the proposed LDV protocol as well as the conventional OBD method. Histograms of the measured $d_{eff}$ amplitudes for both methods are compared in Fig. 4. Not only is $d_{eff}$ heavily overestimated by the OBD method in most cases, but the OBD measurements are also very inconsistent between different cantilevers. Conversely, the LDV measurements result in consistent values of $d_{eff}$. The LDV phase shift histograms (not shown) are also remarkably well-behaved, with the peaks separated by the expected 180° between opposite domains.

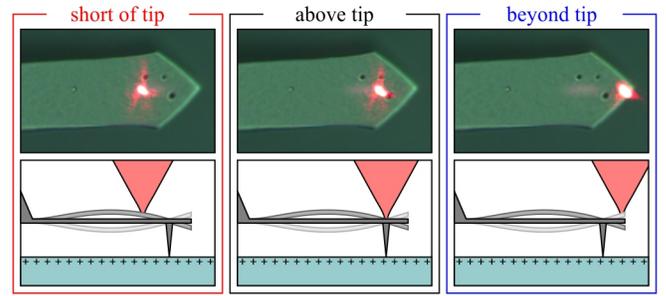

a) three scenarios of laser spot locations

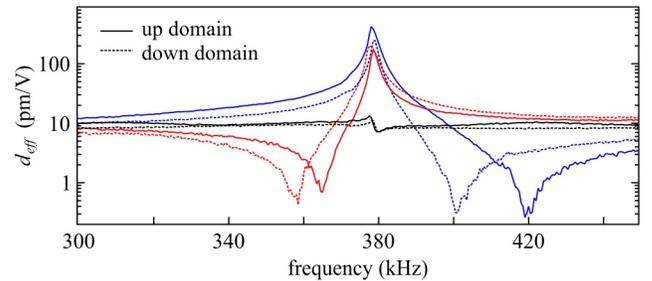

b) contact resonance frequency response

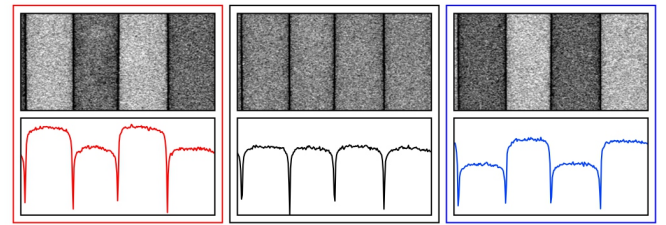

c) $d_{eff}$ images and average traces

Figure 3: (a) Three top-view photographs of laser spot locations are shown, with side-view illustrations. (b) The cantilever frequency response acquired while in contact with the PPNL surface on both negative (down) and positive (up) domains. Different colors correspond to different LDV spot locations on the cantilever using the convention established in (a). The black transfer function has minimal resonant response because the LDV spot is located directly above the tip. (c) Sub-resonance contact images (40 $\mu m$ wide) of $d_{eff}$ taken at 25 kHz with the LDV spot in the three different locations. The graphs below the images are the average of all scan lines in each image. Note that the OBD spot is used only to maintain a constant DC force throughout this experiment.

The LDV measurements consistently provided $d_{eff}$ values near 8.4 pm/V, while the best estimate from bulk characterization of this sample is 27 pm/V. This suggests that, although the metrological AFM accurately measures tip displacement, other sample and cantilever-specific sources of error remain. For example, finite stiffness of the tip-sample contact, boundary clamping effects, non-uniformity of the electric field from the tip and electrical resistance from absorbates or defects at the tip-sample junction of the cantilever can lead to an underestimation of the piezoelectric sensitivity.[24] Indeed, these remaining sources of systematic error can be elucidated in future PFM experiments now that the major issues with repeatability and accuracy of piezoelectric sensitivity measurements have been resolved with use of an integrated LDV.

In addition to electrostatic coupling, there are other sources of background signal that can cause crosstalk in the PFM response. Instrumental electrical resonances may cause a background in the PFM signal. The dangers of a background stray response in the AFM while making PFM measurements have already been elaborated.[25] In the case of the Cypher AFM used here, these effects have been eliminated through careful design of the electrical signal routing and shielding.

A related electromechanical technique that would benefit from combined OBD and LDV measurements is electrochemical strain microscopy[26,27] (ESM). More recently developed than PFM, ESM relies on an oscillating tip-sample bias to induce localized ionic motion, which in turn causes a strain that is coupled to the cantilever through the sharp tip. As with the PFM measurements, metrological AFM measurements of the tip displacement during ESM experiments can provide quantitative measurements of strains induced by ion motion in the sample.

We have developed a metrological AFM that directly and simultaneously measures displacements of the cantilever (or sample) with a LDV rather than inferring it from angular motion of the cantilever measured by OBD. The simultaneous use of both LDV and OBD sensors enables *in-situ* characterization of cantilever dynamics during regular AFM operation, as well as more accurate quantification of local piezoelectric sensitivity – a longstanding goal of nanoscale electromechanics research.

The authors acknowledge useful comments and edits from Donna Hurley.

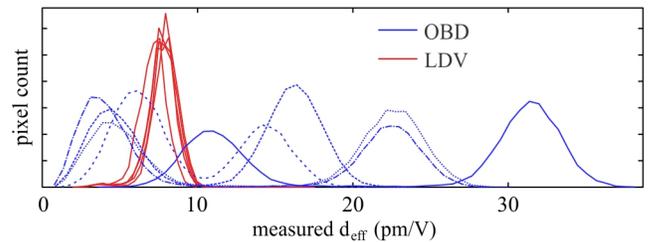

Figure 4: Histograms of the piezoelectric sensitivty $d_{eff}$ for five different cantilevers measured with optical beam deflection (OBD) and with a laser Doppler vibrometer (LDV). The LDV histograms consistently yield values of close to 8.4 pm/V, while the OBD histograms range from 3 to 32 pm/V, demonstrating the irreproducibility of OBD PFM experiments. Note that each OBD measurement has two maxima because the signal differs from up and down domains.

[1] G. Binnig, C.F. Quate, and Ch. Gerber, Phys. Rev. B **56**, 930 (1986).

[2] Martin, Y., C. C. Williams, and H. Kumar Wickramasinghe, *Journal of Applied Physics* 61.10 4723 (1987).

[3] S. Alexander, L. Hellemans, O. Marti, J. Schneir, V. Elings, P.K. Hansma, M. Longmire and J. Gurley, J. Appl. Phys. **65**, 164 (1989).

[4] G. Meyer and N.M. Amer, Appl. Phys. Lett. **53**, 1045 (1988).

[5] M. Bauer, F. Ritter and G. Siegmund, Proc. SPIE 4827, 50 (2002).

[6] P. Güthner and K. Dransfeld, Appl. Phys. Lett. **61**, 1137 (1992).

[7] C. Harnagea, A. Pignolet, M. Alexe, D. Hesse and U. Gösele, Appl. Phys. A: Mater. Sci. Process. **70**, 1 (2000).

[8] H. Bo, Y. Kan, X. Lu, Y. Liu, S. Peng, X. Wang, W. Cai, R. Xue and J. Zhu, J. Appl. Phys. **108**, 042003 (2010).

[9] GHz ferroelectric ref

[10] S. Jesse, B. Mirman and S.V. Kalinin, Appl. Phys. Lett. **89**, 022906 (2006).

[11] C. Harnagea, M. Alexe, D. Hesse and A. Pignolet, Appl. Phys. Lett. **83**, 338 (2003).

[12] I.K. Bdikin, V.V. Shvartsman, S.H. Kim, J.M. Herrero and A.L. Kholkin, Mat. Res. Soc. Symp. Proc. **784**, C11.3 (2004).

[13] B.D. Huey, in *Nanoscale Phenomena in Ferroelectric Thin Films*, ed. S. Hong (Kluwer, New York, 2004), pp. 239-262.

[14] B.J Rodriguez, C. Callahan, S.V Kalinin and R. Proksch, Nanotechnology **18**, 475504 (2007).

[15] S. Jesse, S.V. Kalinin, R. Proksch, A.P. Baddorf and B.J. Rodriguez, Nanotechnology **18**, 435503 (2007).

[16] K. Franke, H. Huelz and M. Weihnacht, Surf. Sci. **415**, 178 (1998).

[17] S. Hong, J. Woo, H. Shin, J. U. Jeon, Y.E. Park, E. Colla, N. Setter, E. Kim and K. No, J. Appl. Phys. **89**, 1377 (2001).

[18] C. Harnagea, M. Alexe, D. Hesse and A. Pignolet, Appl.Phys. Lett. **83**, 338 (2003).

[19] B.D. Huey, C. Ramanujan, M. Bobji, J. Blendell, G. White, R. Szoszkiewicz and A. Kulik, J. Electroceram. **13**, 287 (2004).

[20] AR-PPLN LiNbO$_3$ test sample, Asylum Research, Santa Barbara, CA: http://www.asylumresearch.com/Products/AR-PPLN/AR-PPLN.shtml .

[21] J.W. Burgess, J. Phys. D **8**, 283 (1975).

[22] H.-N. Lin, S.-H. Chen, S.-T. Ho, P.-R. Chen, and I.-N. Lin, J. Vac. Sci. Technol. B **21**, 916 (2003).

[23] R. Proksch, http://arxiv.org/abs/1409.0133 and J. App. Phys., in Press (2015).

[24] T. Jungk, Tobias, Ákos Hoffmann, and Elisabeth Soergel, *Applied Physics A* 86.3 353 (2007).

[25] T. Jungk, A. Hoffmann, and E. Soergel, Appl. Phys. Lett., **89**, 163507 (2006).

[26] N. Balke, S. Jesse, A.N. Morozovska, E. Eliseev, D.W. Chung, Y. Kim, L. Adamczyk, R.E. Garcıa, N. Dudney and S.V. Kalinin, Nature Nanotech. **5**, 749 (2010).

[27] N. Balke, S. Jesse, Y. Kim, L. Adamczyk, A. Tselev, I.N. Ivanov, N.J. Dudney and S.V. Kalinin, Nano Lett. **10**, 3420 (2010).